\documentclass{aastex631} 

\usepackage{color}
\usepackage{bm}
\usepackage{ amssymb, amsmath }

\newcommand{\g}{\textbf}

\newcommand{\BB}{\bm{B} }

\newcommand{\VV}{\bm{V} }
\newcommand{\YY}{\bm{Y} }

\newcommand{\bb}{\bm{b} }

\newcommand{\vv}{\bm{v} }
\newcommand{\xx}{\bm{x} }

\newcommand{\zz}{\bm{z} }

\renewcommand{\bell}{\boldsymbol{\ell}}

\shorttitle{3D energy cascade rate and Yaglom flux in the Earth's magnetosheath}
\shortauthors{Pecora F.}
\graphicspath{{./}}

\begin{document}

\title{In-situ observations of the three-dimensional energy cascade rate and Yaglom flux in the Earth's magnetosheath}

\author[0000-0003-4168-590X]{Francesco Pecora}
\affiliation{Department of Physics and Astronomy, University of Delaware, Newark, DE 19716, USA}
\email{fpecora@udel.edu}



\begin{abstract}

Measuring the energy cascade rate in space plasmas is a challenging task for several reasons. This quantity is (i) inherently three-dimensional (ii) scale-dependent, (iii) anisotropic in the interplanetary plasma, and (iv) requires measurements of plasma parameters in at least four points. Here, we show how three of such problems have been addressed by applying the novel Lag Polyhedra Derivative Ensemble (LPDE) technique to the Magnetospheric Multiscale (MMS) mission in the Earth's magnetosheath.

\end{abstract}



\section{Introduction}
\label{sec:intro}

%

One of the hardest quantities to measure in turbulent space plasma is the energy transfer rate. This quantity determines how fast the energy is transferred across scales from the large energy reservoirs that feed the system to the scales where nonlinear interactions dominate, and where energy is eventually dissipated. The energy transfer (or cascade or dissipation) rate is inherently three-dimensional and, in general, the assumption of anisotropy cannot be made. Currently, only the Magnetospheric Multiscale (MMS) mission \citep{burch2016MMS} provides multipoint measurements that can be used to effectively estimate such quantity without significant levels of approximations.

Monumental theoretical efforts have been devoted to the description of the energy evolution in turbulent media starting from hydrodynamics \citep{karman1938statistical}. The von K\'arman-Howarth (vKH) equation derived in the homogeneous incompressible hydrodynamic context describes a balance across scales between the temporal variations, nonlinear transfer, and dissipation. Eventually, these results were extended to homogeneous incompressible magnetohydrodynamics (MHD) \citep[][hereafter PP]{politano1998dynamical_PP1, Politano1998vonkarman_PP2}. Of relevance in these works are the two presented expressions for anisotropic and isotropic magnetized fluids. The second of these has found great application in space plasmas \citep{macbride2005turbulence,sorrisovalvo2007observation, stawarz2009turbulent, marino2011magnetohydrodynamic, bandyopadhyay2018incompressive, andres2022incompressible}. Whereas, the former has been considerably less studied in space plasmas \citep{macbride2008turbulent,osman2011anisotropic,pecora2023threedimensional_PRL}.

The complexity of plasmas called for extensions of PP work to realms different from incompressible MHD. A number of generalizations to PP have been developed including compressibility effects in the description of the energy transfer. These originated numerous observational works \citep{carbone2009scaling, galtier2011exact, banerjee2013exact, hadid2017energy, andres2017alternative, andres2018energy, andres2021evolution}. Additionally, the Hall effect was also included in the description of incompressible \citep{galtier2008vonkarman, hellinger2018vonkarman_HMHD, hellinger2021spectral} and compressible \citep{andres2018exact} MHD, as well as to kinetic particle-in-cell reconnection simulations \citep{adhikari2023effect}. For a general review, see \citet{marino2023scaling}.

Anisotropy effects are crucial in determining different behavior of the turbulent cascade in different directions with respect to a preferred direction. This was first observed in hydrodynamics experiments where the anisotropy was induced by spinning the system \citep{lamriben2011direct}, and later in MHD simulations where anisotropy is generated by a mean magnetic field \citep{verdini2015anisotropy,wang2022strategies,jiang2023energy}. Experiments and simulations show that the isotropic estimates of the cascade rate show great variability, and averaging over numerous directions is needed in order to obtain an accurate evaluation of the energy cascade rate from unidirectional samplings \citep{nie1999note}.

Previous works started exploring the possibility of multidirectional estimates of the energy cascade rate through directional averages \citep{osman2011anisotropic,bandyopadhyay2018incompressive}, or a combination of 2D+1D models \citep{macbride2008turbulent,stawarz2009turbulent}. \citet{stawarz2009turbulent} also noted that the convergence of the third-order structure function requires several months' worth of data which is also about the same timespan needed to obtain enough directions for an axisymmetric estimate of the cascade rate \citep{osman2011anisotropic}. This, of course, rules out the possibility of using such approaches for local analyses.

Recently, a novel technique, Lag Polyhedra Derivative Ensemble (LPDE), was developed to provide fully three-dimensional statistically significant estimates for the energy cascade rate provided that multipoint measurements are available. The technique was tested in simulations \citep{pecora2023helioswarm} and applied to Magnetospheric Multiscale \citep[MMS,][]{burch2016MMS} data in the magnetosheath \citep{pecora2023threedimensional_PRL}. In the present paper, additional measurements of the energy cascade rate and observations of the Yaglom flux (that is responsible for the nonlinear transfer of energy across scales) are reported. The paper is organized as follows, in Section~\ref{sec:theo} the theoretical framework of the von~K\'arm\'an-Howarth (vKH) equation is presented, in Section~\ref{sec:data} the MMS data used for the analyses is described, the LPDE technique is illustrated in Section~\ref{sec:lpde}, results are shown in Section~\ref{sec:res}. Results are discussed and conclusions are drawn in the final Section.

\section{Theoretical framework}
\label{sec:theo}

The extension of the von~K\'arm\'an-Howarth equation \citep{karman1938statistical} to MHD carried out by \citet{politano1998dynamical_PP1, Politano1998vonkarman_PP2} involves increments of the Els\"asser fields $\zz^\pm = \vv \pm \bb$, where $\vv$ and $\bb$ are the velocity and the magnetic fields, the latter is in Alfv\'en units -- normalized to $\sqrt{4 \pi n_p m_p}$ for proton number density $n_p$ and mass $m_p$. The full equations read:

\begin{equation}
    \displaystyle\frac{\partial}{\partial t}\langle |\delta \mathbf{z}^\pm|^2 \rangle + \mathbf{\nabla}_{\boldsymbol{\ell}} \cdot \langle \delta \mathbf{z}^\mp |\delta \mathbf{z}^\pm|^2 \rangle - 2 \nu \nabla^2_{\boldsymbol{\ell}}\langle |\delta \mathbf{z}^\pm|^2 \rangle = - 4\epsilon^\pm
    \label{eq:vKH}
\end{equation}

where $\nu$ is the kinematic viscosity. The Els\"asser increments are defined at vector lag  $\bell$ as $\delta\zz^\pm(\bell) = \zz^\pm(\xx) - \zz^\pm(\xx + \bell)$. Derivatives $\mathbf{\nabla}_{\boldsymbol{\ell}}$ are in lag (or increment) space and $\epsilon = \frac{\epsilon^+ + \epsilon^-}{2}$ is the energy transfer rate. The three terms on the left-hand side can be labeled as T, N, and D respectively, and will now be described.

The T term represents the temporal variation of the energy. Because of its nature, it cannot be measured with spacecraft since it would require the instruments to follow the same parcel of plasma in time. Indeed, the usually employed Taylor hypothesis \citep{taylor1938spectrum_frozenin} states that temporal variations can be considered negligible within an appropriate observational time window, and fluctuations are representative of spatial structures. 

The dissipative effects are included in the D term. Aside from the computation of the Laplacian, evaluating this term requires knowledge of the viscosity $\nu$. This is a separate problem for noncollisional plasmas. Recent works started shedding light on possible proxies for the dissipation function \citep{yang2022pressure, pezzi2021dissipation, pezzi2021current, yang2024effective}.

Finally, the transfer of energy across scales due to the nonlinear interactions is described by the term N. It includes a mixed third-order structure function $\delta \mathbf{z}^\mp |\delta \mathbf{z}^\pm|^2$ that, in analogy with Yaglom’s work \citep{yaglom1949local}, is referred to as Yaglom flux. The divergence, in lag space, of the Yaglom flux, is the main ingredient evaluated in space plasmas because of the difficulties the other two present.

The different natures of the T, N, and D terms are such that they are dominant in separate ranges of scales in turbulent systems with a large enough Reynolds number. Their sum, by virtue of Eq.~\ref{eq:vKH}, is always proportional to the energy transfer rate. When the range of scales pertains to the inertial range of turbulence, ideally, the T and D terms are negligible, and therefore the cascade rate can uniquely be determined by the N term. A similar argument holds at large scales (beyond the correlation or integral scale) for the T term and at small scales for D. When in the inertial range and T and D are negligible, the vKH equation reduces to the so-called Yaglom's law:

\begin{equation}
    \mathbf{\nabla}_{\boldsymbol{\ell}} \cdot \YY = - 4\epsilon^\pm,
    \label{eq:yaglom}
\end{equation}

\noindent where $\YY = \langle \delta \mathbf{z}^\mp |\delta \mathbf{z}^\pm|^2 \rangle $ is the Yaglom flux. However, if T and D are not negligible, the solution to Eq.~\ref{eq:yaglom} gives a partial transfer rate from which the contributions from the other terms are missing. Therefore, it can be interpreted as a lower threshold for the total transfer rate. This scenario is representative of the analyses proposed in these papers since the separations between the MMS spacecraft are such that they are smaller than the scales within the inertial range as delimited by the ion skin depth and the correlation length (see below).

\section{MMS data}
\label{sec:data}

All data used are obtained from the MMS mission \citep{burch2016MMS}. The magnetic field measurements are provided by the Fluxgate magnetometers \cite[FGM,][]{russell2016magnetospheric}, proton velocity and electron density from the Fast Plasma Investigation \citep[FPI,][]{pollock2016fast}. Electron density has been used instead of ion density since it is generally more accurate, and quasi-neutrality is assumed. Different products have different resolutions in burst mode. The magnetic field is available at 128~Hz, ion velocity at 150~ms, and electron density at 30~ms. Velocities are despun using the appropriate product provided by the mission. Then, all quantities are re-sampled at the common frequency of 150~ms. Density measurements  larger than 50~cm$^{-3}$ are discarded since they can be polluted by instrumental inaccuracies. In the intervals listed below, MMS was in the magnetosheath. To characterize these intervals, relevant quantities have been computed and reported in Table~\ref{tab:table1}. Listed are the magnetic and mass density fluctuations defined as the root mean square normalized by the mean, the plasma beta (ratio between thermal and magnetic pressure), the correlation time and length, and the ion inertial length. The correlation time is obtained as the $1/e$-folding of the normalized correlation function for the magnetic field defined as $\tilde{R} = R(\tau)/R(0) = \langle \BB(t) \cdot \BB(t+\tau) \rangle_t / \langle |\BB(t)|^2 \rangle_t $ \citep{matthaeus1999correlation}. The correlation length $\lambda_c$ is obtained by multiplying the correlation time by the average solar wind speed of the considered interval.


\begin{table*}[ht]
    \centering
    \begin{tabular}{ccccccccc}
        & Date (UTC) & $\delta b / B$ & $\delta \rho / \rho $ & $\beta$ & $\tau_c$(s) & $\lambda_c$(km) & $d_i$(km)  \\
        \hline
        I   & 2017 Sep 28 06:31:33 -- 07:01:43 & 0.51 & 0.19 & 5.9  & 81.0 & 41229 & 46  \\
        II  & 2017 Nov 10 22:35:43 -- 22:52:03 & 3.17 & 0.43 & 8.3  & 2.9  & 1377 & 74  \\
        III & 2017 Dec 21 07:21:54 -- 07:48:01 & 1.92 & 0.31 & 4.7  & 6.1  & 652 & 50  \\
        IV  & 2017 Dec 26 06:12:43 -- 06:52:23 & 0.82 & 0.21 & 4.5  & 16.3 & 3712 & 48  \\
        V   & 2018 Apr 19 05:10:23 -- 05:41:53 & 2.99 & 0.29 & 15.0 & 5.1 & 1168 & 36 
    \end{tabular}
    \caption{Indicated are the dates and times of the analyzed intervals, magnetic $\delta b/B$ and density $\delta \rho/\rho$ fluctuation levels (root mean square over mean), proton plasma beta $\beta$ (ratio between thermal and magnetic pressure), correlation time $\tau_c$ and length $\lambda_c$, ion inertial length $d_i$. Table adapted from \protect\cite{pecora2023threedimensional_PRL}.}
    \label{tab:table1}
\end{table*}

\section{LPDE}
\label{sec:lpde}

The LPDE technique has been first described and applied to MHD simulations in \citep{pecora2023helioswarm} and later extended to include tetrahedra quality checks and applied to MMS data in \citep{pecora2023threedimensional_PRL}. LPDE exploits the nature of Eq.~\ref{eq:yaglom} that requires derivatives to be computed in lag space in order to obtain numerous estimates of the energy transfer. Through this technique, it is possible to obtain a statistically significant number of estimates for the (partial) energy transfer rate ranging from several hundreds for MMS-like configurations up to $\sim 10^6$ for the HelioSwarm \citep{spence2019helioswarm, klein2023helioswarm} nine-spacecraft constellation. Below, the main points of LPDE are described.

The main ingredient of Eq.~\ref{eq:yaglom} is the Yaglom flux $\YY$ that is defined through increments of the Els\"asser fields $\delta \zz^\pm = \zz^\pm(\xx+\bell) - \zz^\pm(\xx)$ (in the following, the $\pm$ superscript is dropped when unnecessary). Such increments can be computed in several different ways in the realm of spacecraft time series. If the Taylor hypothesis is used \citep{taylor1938spectrum_frozenin}, the position $\xx$ can be obtained as $-\VV_{sw} t$, where $\VV_{sw}$ is the solar wind speed, and $t$ is the observational time. At the price of neglecting temporal variations, this approach allows the computation of increments with single-spacecraft measurements. The other approach requires multi-spacecraft measurements e.g. a pair of spacecraft $i,j$, at positions $\xx_i$ and $\xx_j$ respectively such that $\xx_{ij} = \xx_i - \xx_j \equiv \bell$. MMS has a constellation of four spacecraft and therefore a value of the Yaglom flux for each of the six vector baselines $\delta\zz_{ij}(\xx_{ij})=\zz_i(\xx_i)-\zz_j(\xx_j)$, where $i,j=1,\dots 4,~\;~i<j$. However, the Yaglom flux, being a (mixed) third-order structure function, is an odd function of the vector lag. Therefore, one has knowledge of $\YY$ at $\xx_{ij}$ and $\xx_{ji}$, namely $\YY(\xx_{ij}) = -\YY(\xx_{ji})$. This increases the available measurements from 6 to 12. This may not seem like a great increase, however, the technique is based on combinations and the total estimates of the energy transfer rates depend on $C_{N}^{K}=\binom{N}{K}=\frac{N!}{K!(N-K)!}$, where $N$ is the number of available points (the values of $\YY$), and $K$ is the number of elements in the subsets they are arranged in. To use curlometer-like techniques, $K=4$. Then, it is immediate to recognize that the pool of estimates for the energy transfer rates boosts from $C_6^4=15$ to $C_{12}^4=495$.

Now, a correction has to be made to take into account duplicate sets of points. For example, the same $\epsilon$ will be obtained if the divergence is computed over some 4 points or over the reflections of those same 4 points because this operation builds the same tetrahedron with the same values of $\YY$ at its vertices. An additional correction can be made by excluding those tetrahedra whose barycenter coincides with the origin (zero lag) that is of dubious physical interpretation. Such tetrahedra are those built with any two points and their reflections. Their number is $C_6^2 = 15$. Finally, the total number of independent and physically acceptable estimates is $\frac{1}{2}( C_{12}^{4} - C_{6}^{2} ) = 240$ which is still much larger than $15$ and also statistically significant.

Additional constraints to improve the quality of such estimates can be included by measuring the elongation (E) and planarity (P) parameters for the tetrahedra used to solve Eq.~\ref{eq:yaglom} \citep{paschmann1998analysis}. Elongation and planarity are defined as $E=1-\sqrt{\lambda_2/\lambda_1}$ and $P=1-\sqrt{\lambda_3/\lambda_2}$ where $\lambda_s$ are the eigenvalues of the volumetric tensor $L_{ij} = \frac{1}{K}\sum_{\alpha=1}^N(\ell_{\alpha i}\ell_{\alpha j}-\ell_{b i}\ell_{b j} )$ where $\ell_{\alpha i}$ is the i-th component of the vertex $\alpha$ associated with each tetrahedron, and $K=4$. $\ell_{b j}=\langle\ell_{\alpha j}\rangle_\alpha$ is the position of the barycenter of the considered tetrahedron with vertices $\alpha$. Well-behaved tetrahedra are those whose E and P values are closer to the origin of the EP plane. These are the most regular tetrahedra and provide the most reliable estimates. If the position in the EP plane is defined as $d_{EP} = \sqrt{E^2 + P^2}$, the tetrahedra that provide estimates with error potentially smaller than $20\%$ are those such that $d_{EP}<0.6$ \citep[see page 408 of][]{paschmann1998analysis}.

\section{Results}
\label{sec:res}

\begin{figure*}[ht]
    \centering
    \includegraphics[width=0.23\textwidth]{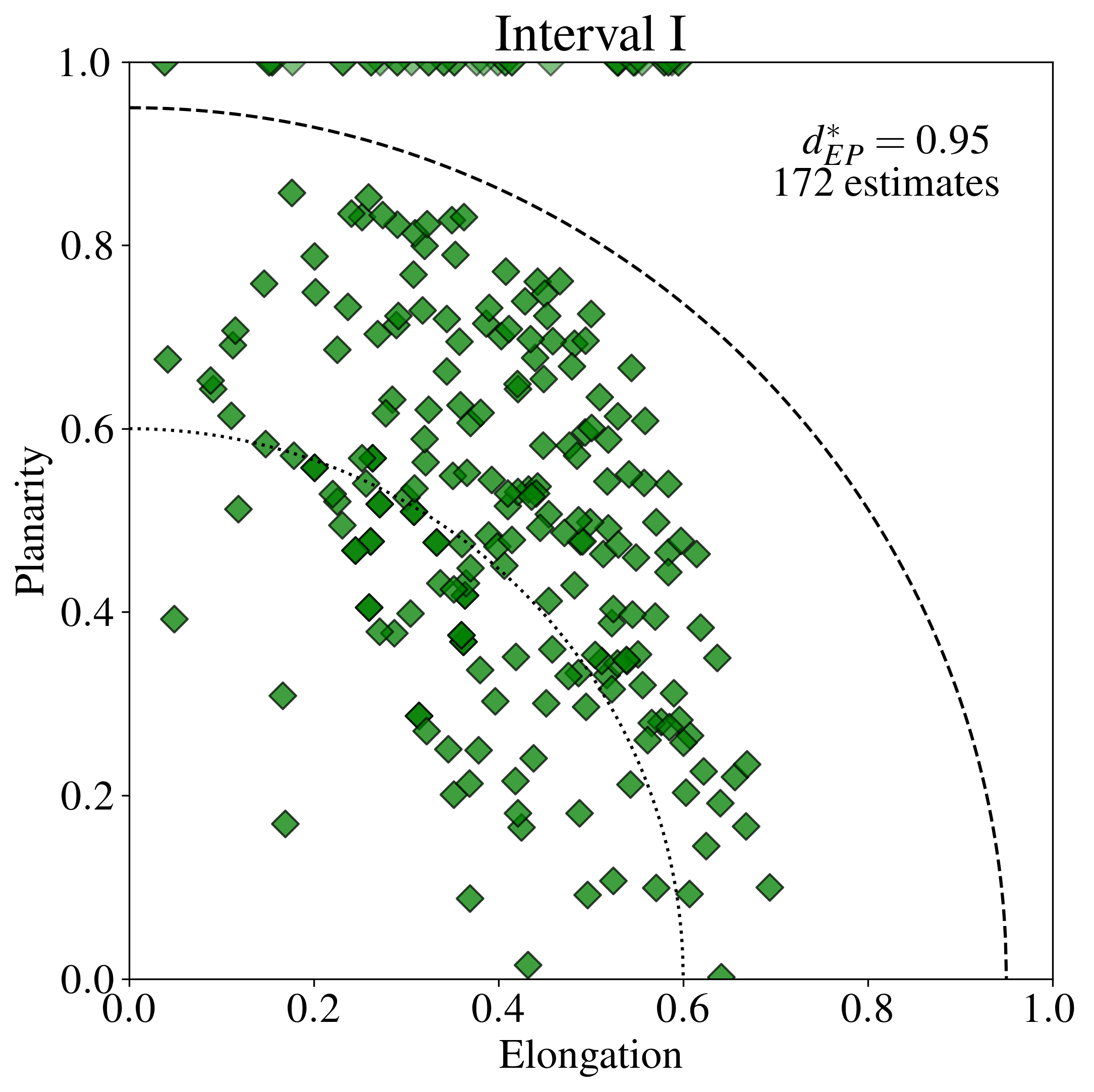}
    \includegraphics[width=0.23\textwidth]{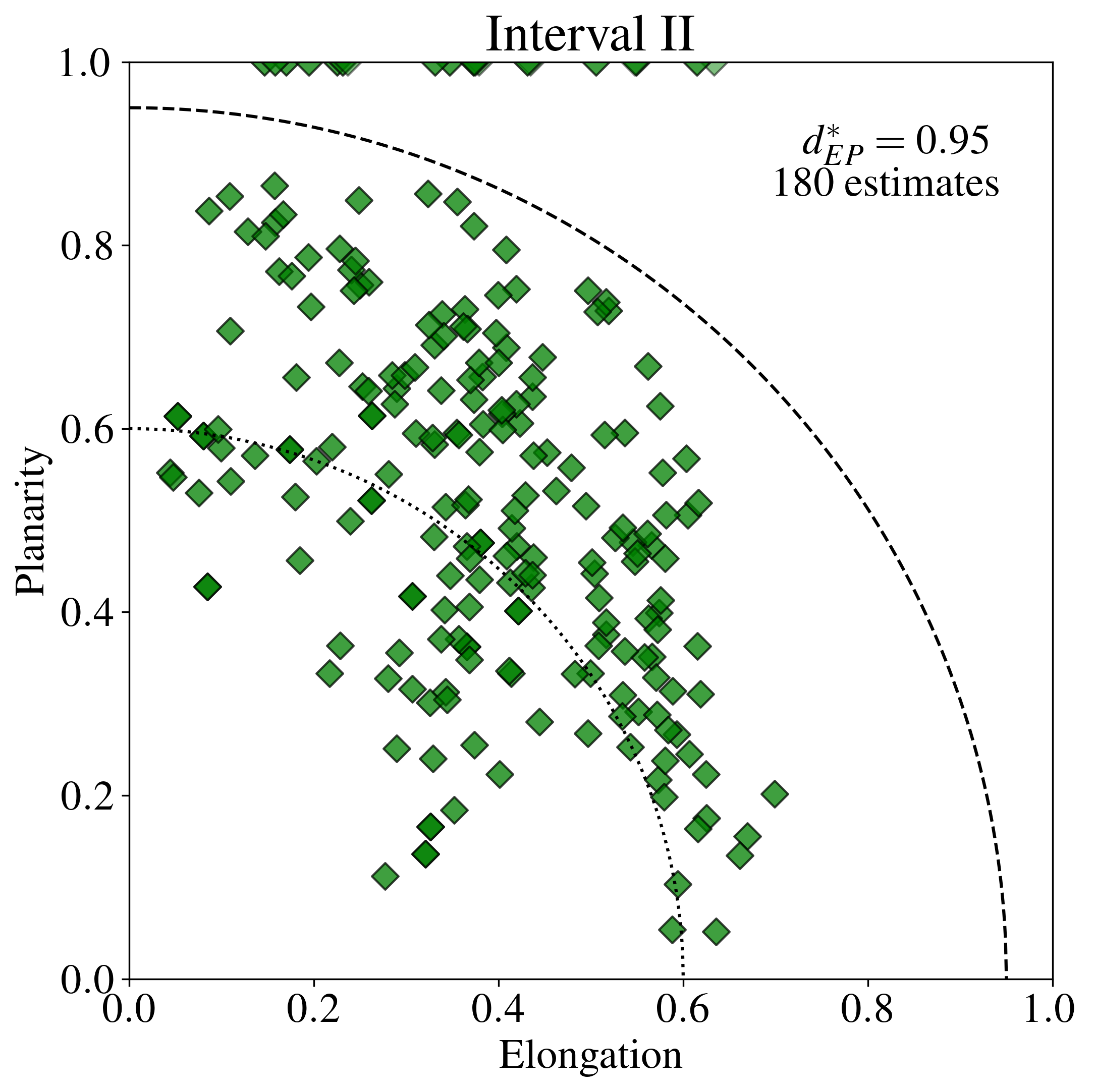}
    \includegraphics[width=0.23\textwidth]{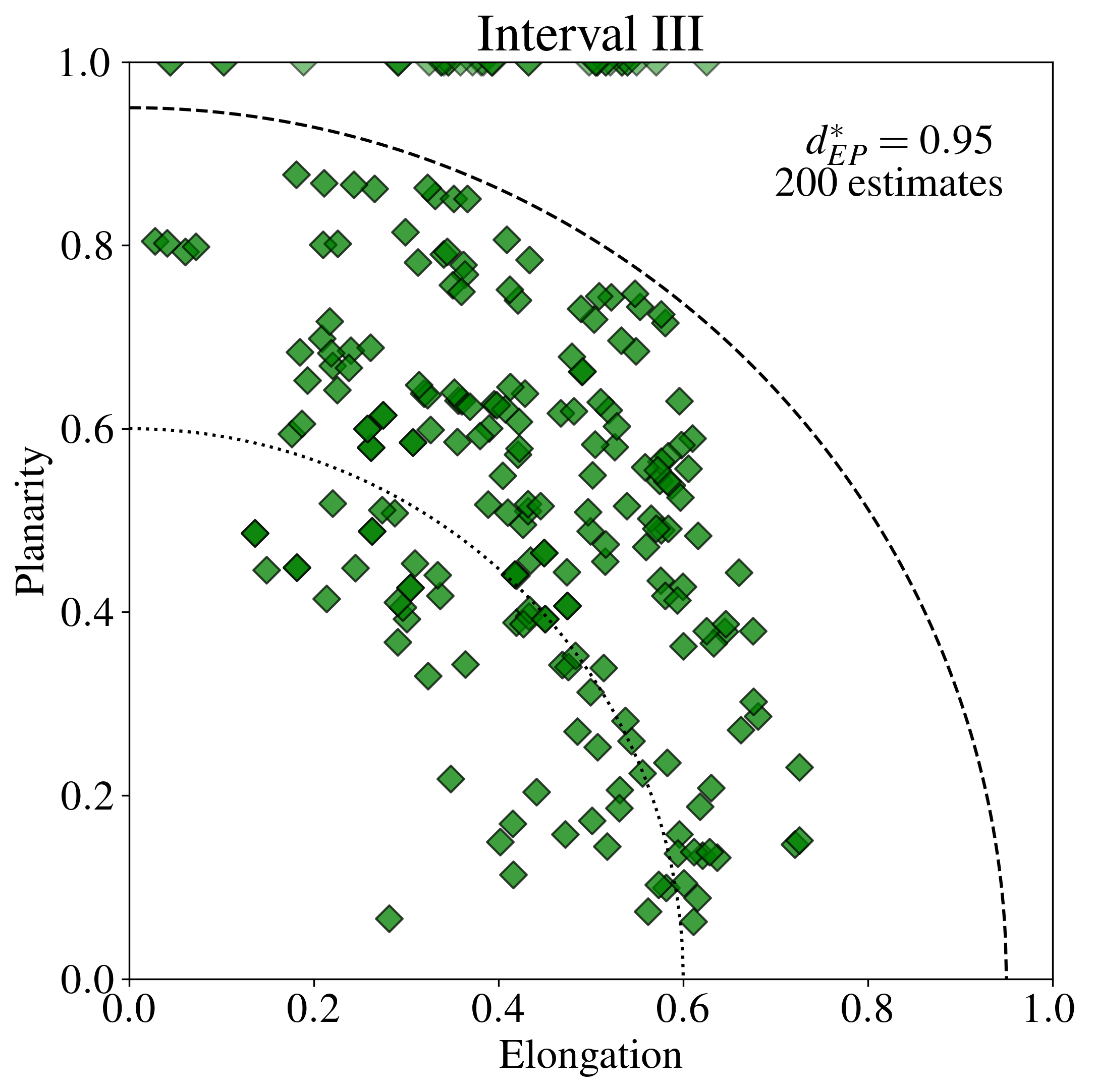}
    \includegraphics[width=0.23\textwidth]{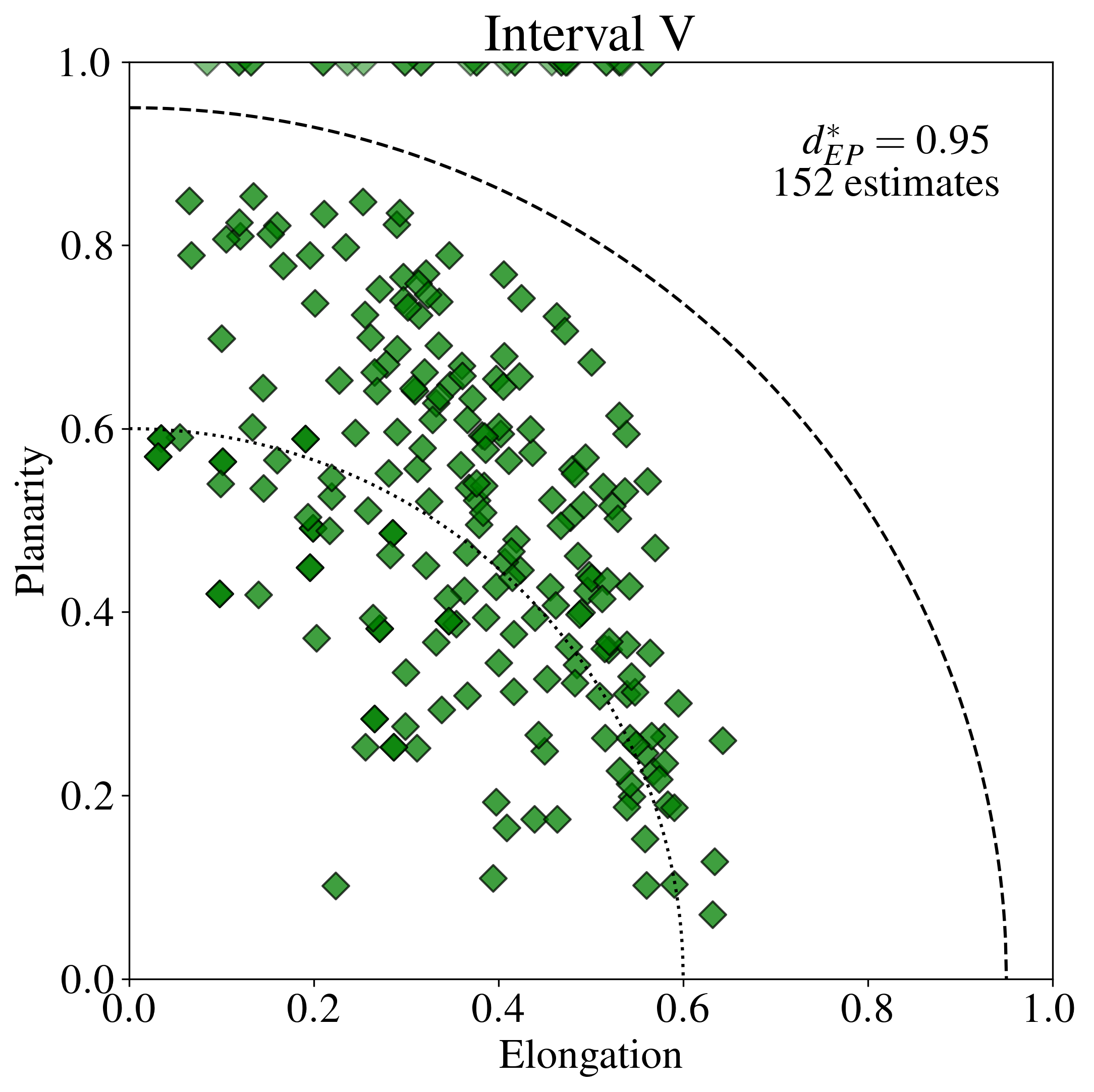}
    \caption{Elongation (E) and planarity (P) parameters for the tetrahedra in lag-space used to solve Yaglom's law. Indicated are the $0.6$ threshold (dotted line) which should be associated with errors in the estimates smaller than $20\%$ and the largest threshold (dashed line) that maintains the same average value for the cascade rate but keeps the most number of points for statistical reasons. Within each panel, the number of points below the relaxed threshold of $0.95$ is also indicated.}
    \label{fig:EP}
\end{figure*}

In this Section, results for intervals I, II, III, and V of Table~\ref{tab:table1} are reported. Interval IV is described in \citep{pecora2023threedimensional_PRL}. Figure~\ref{fig:EP} shows the EP values for the lag-space tetrahedra during the four considered intervals. The figure also reports the initial ``conservative'' $(d_{EP}=0.6)$ and ``relaxed'' threshold alongside the number of estimates that are used to obtain the cascade rate values. The stricter threshold has been increased gradually as long as the average value of the energy transfer rate does not change. This has been done as a convergence test and in order to keep the largest possible number of estimates for statistical reasons.

\begin{figure*}[ht]
    \centering
    \includegraphics[width=0.23\textwidth]{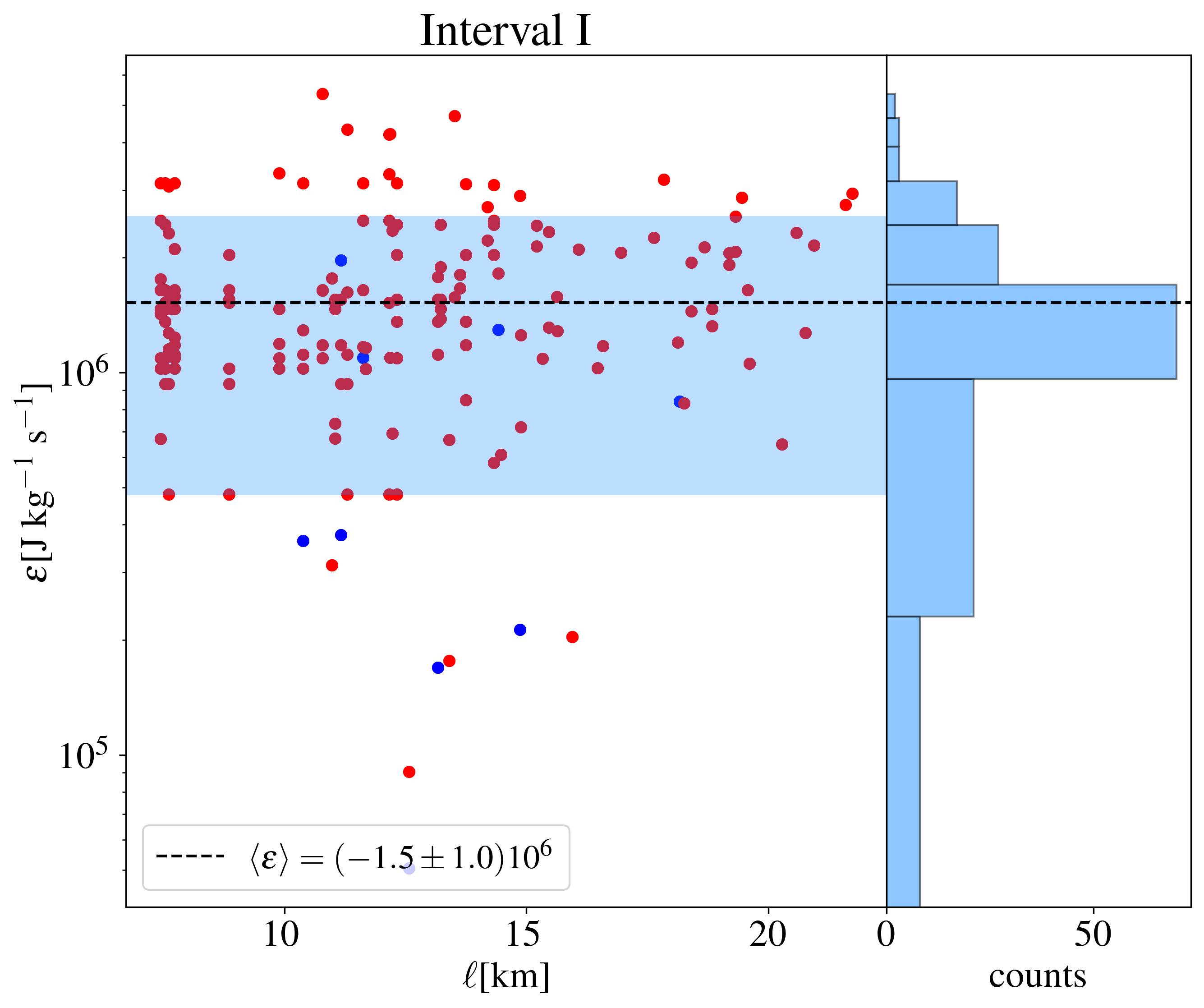}
    \includegraphics[width=0.23\textwidth]{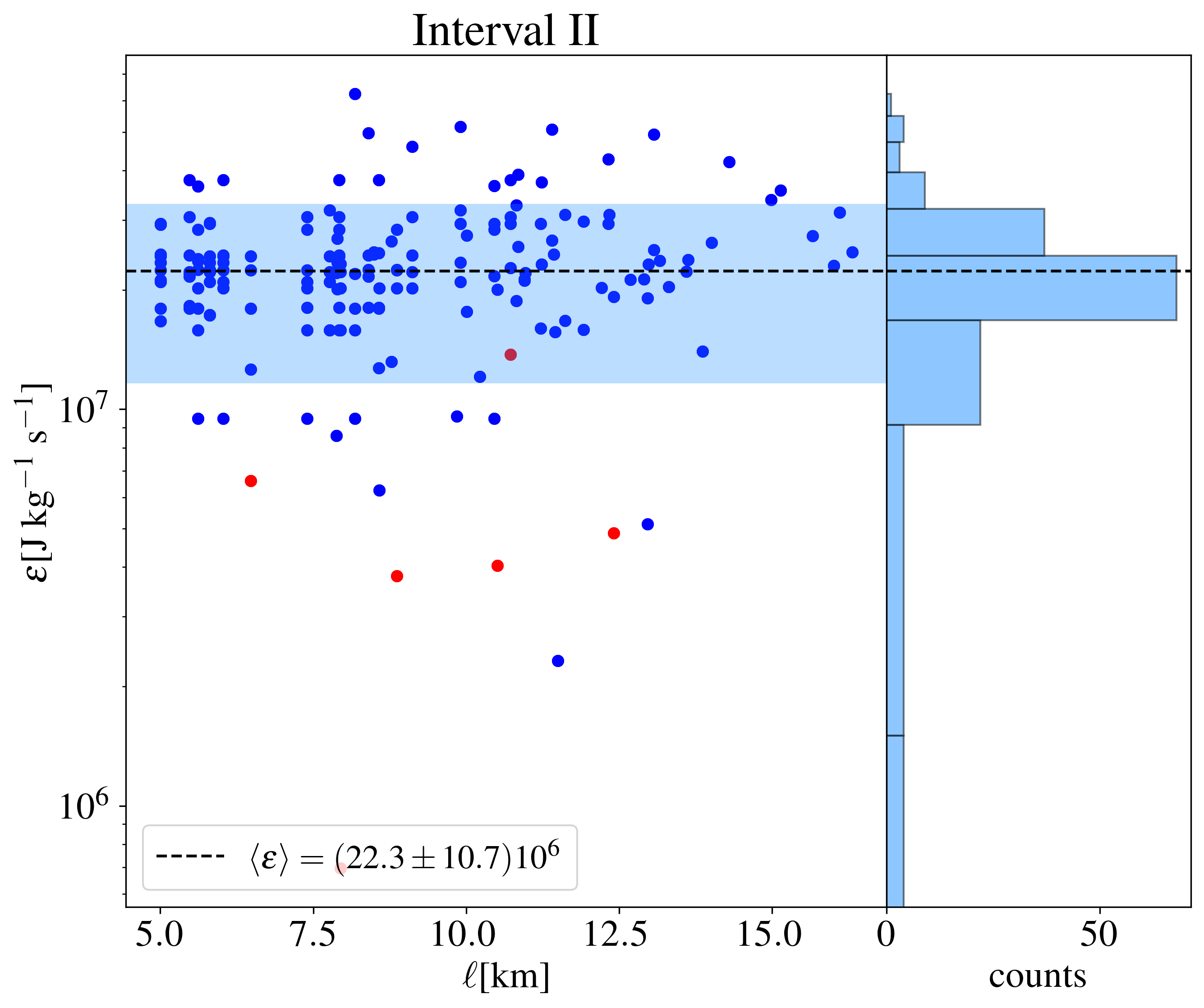}
    \includegraphics[width=0.23\textwidth]{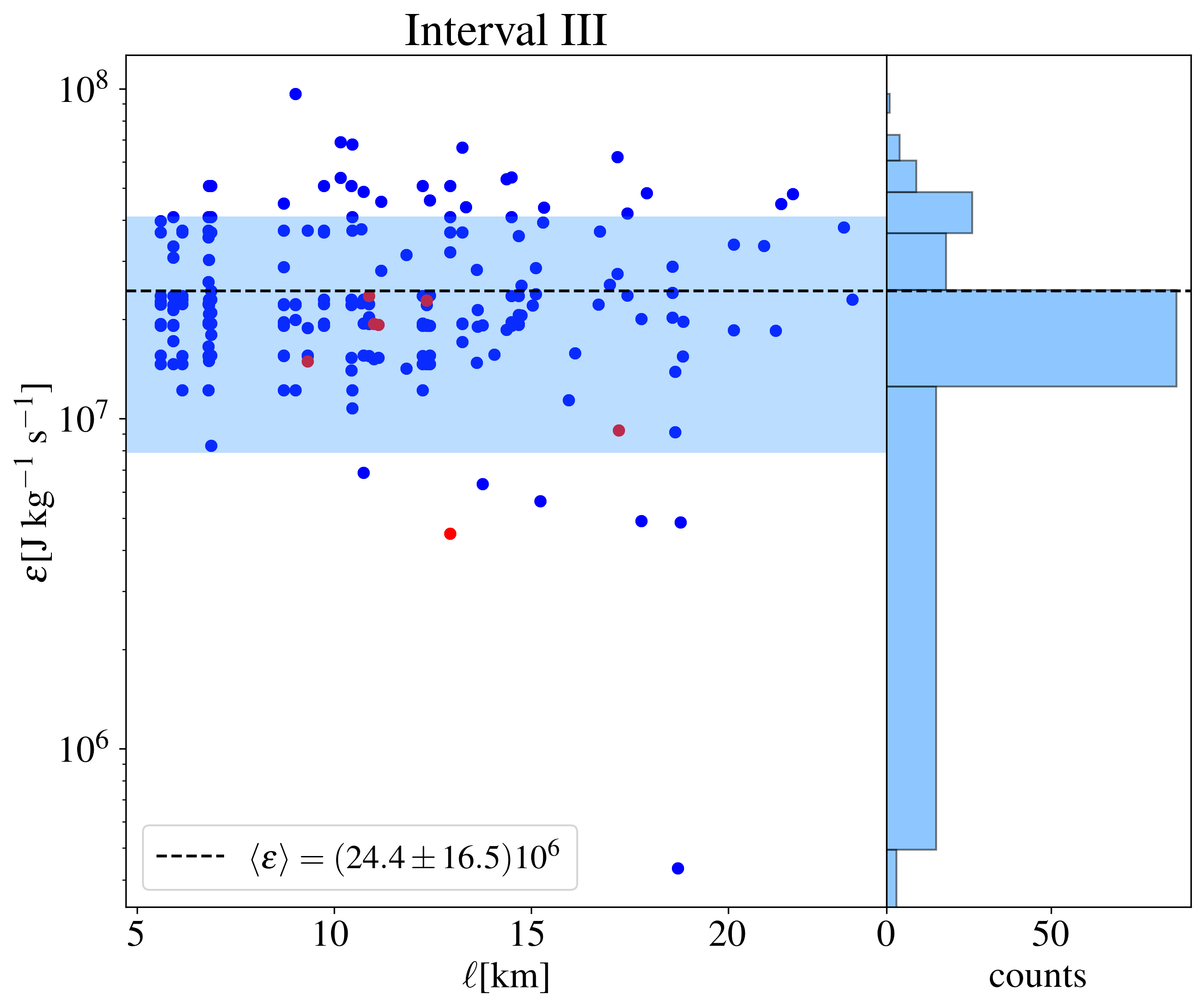}
    \includegraphics[width=0.23\textwidth]{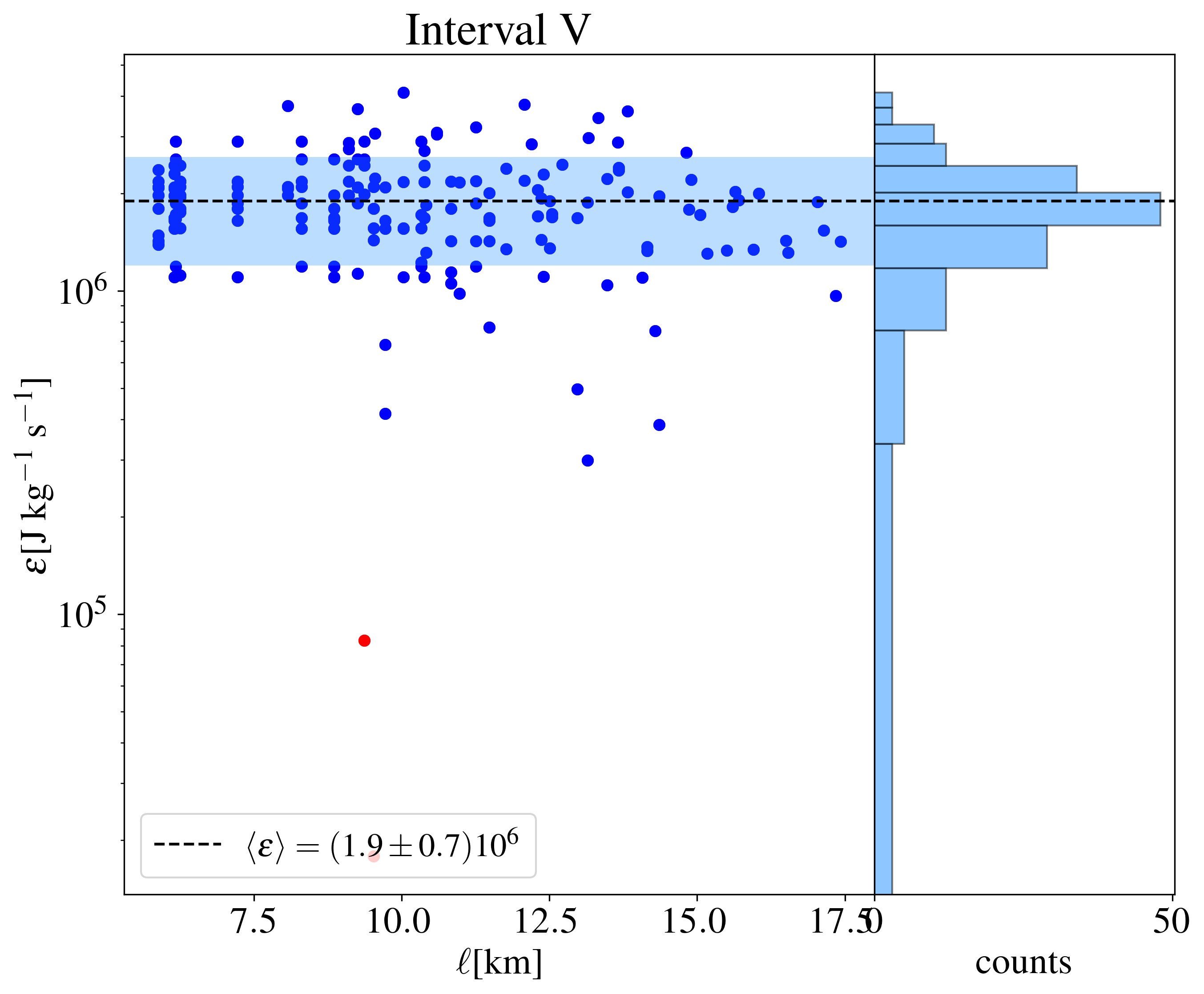}
    \caption{Values obtained for the energy transfer rate for the four analyzed intervals. Red dots indicate negative values. The average (horizontal dashed line) and the variance of the histogram on the right of each figure (light-blue shaded region) are computed using all signed estimates (no absolute values). Notice that each interval is ``polluted'' by a few values of the opposite sign of the majority. However, the large number of estimates provides significant stability to the obtained cascade rates, as also shown by the convergence test in Table~\ref{tab:table2}.}
    \label{fig:eps_hist}
\end{figure*}

The number of points falling within the prescribed threshold reflects the number of tetrahedra available to solve Eq.~\ref{eq:yaglom}. This means that $\sim 150$ to $200$ values of the energy cascade rate are available per single interval granting solid statistical significance to this measure. Figure~\ref{fig:eps_hist} shows all the estimated energy transfer rates including both positive and negative values. The presence of solutions with the opposite sign within a dominantly positive or negative average solution suggests the need for a large number of estimates to accurately measure the cascade rate. A single negative estimate could be misinterpreted as, for example, an inverse cascade if most of the positive estimates are inaccessible. Therefore, having a larger sample size allows for considering the signed value of this quantity with more confidence \citep{hadid2018compressible}.

The average values of the energy transfer rate obtained from LPDE along with the uncertainty obtained from the standard deviation of their histograms are shown in Fig.~\ref{fig:eps_hist} and reported in Table~\ref{tab:table2}. Each row of the table indicates the average value of the energy transfer rate with the associated uncertainty for a certain threshold $d^*_{EP}$ together with the total number N of estimates that fall within the threshold. By increasing the threshold, less ``well-behaved'' tetrahedra are included in the global estimate. However, the large number of available points makes the solution stable with respect to fluctuations induced by less accurate or opposite-sign solutions.

\begin{table*}[ht]
    \centering
    \begin{tabular}{c|cc|cc|cc|cc}
        & \multicolumn{2}{c}{I} & \multicolumn{2}{c}{II} & \multicolumn{2}{c}{III} & \multicolumn{2}{c}{V}  \\
    \hline
      $d^*_{EP}$ &    $\epsilon$  & N   &    $\epsilon$   & N   &  $\epsilon$      & N   &     $\epsilon$ & N \\
    \hline
        $0.6$    & $-1.6 \pm 0.7$ & 50  &  $22.4 \pm 4.7$ & 56  &   $23.8 \pm 10$  & 55  &  $1.9 \pm 0.4$ & 62                 \\
        $0.7$    & $-1.6 \pm 0.7$ & 111 &  $23.4 \pm 7.0$ & 115 &   $23.7 \pm 9.9$ & 98  &  $1.9 \pm 0.5$ & 98  \\
        $0.8$    & $-1.5 \pm 0.8$ & 140 &  $22.3 \pm 8.5$ & 150 &   $23.6 \pm 12$  & 158 &  $1.9 \pm 0.6$ & 126  \\
        $0.95$   & $-1.5 \pm 1.0$ & 172 &  $22.3 \pm 11$  & 180 &   $24.4 \pm 17$  & 200 &  $1.9 \pm 0.7$ & 152  \\
    \hline
    
    \end{tabular}
    \caption{For the intervals I, II, III, and V described in Table~\ref{tab:table1}, reported are the average energy transfer rate (in units of $10^6$~J kg$^{-1}$ s$^{-1}$) with the associated uncertainty measured as the standard deviation of the histograms, and the number N of total available estimates associated with the threshold $d^*_{EP}$. Notice the stability of the evaluation of the energy transfer rate even when less ideal tetrahedra (larger $d^*_{EP}$) are included. The only appreciable effect is the uncertainty becoming larger, as one would expect.}
    \label{tab:table2}
\end{table*}

\subsection{Yaglom flux}

\begin{figure*}[ht]
    \centering
    \includegraphics[width=0.23\textwidth]{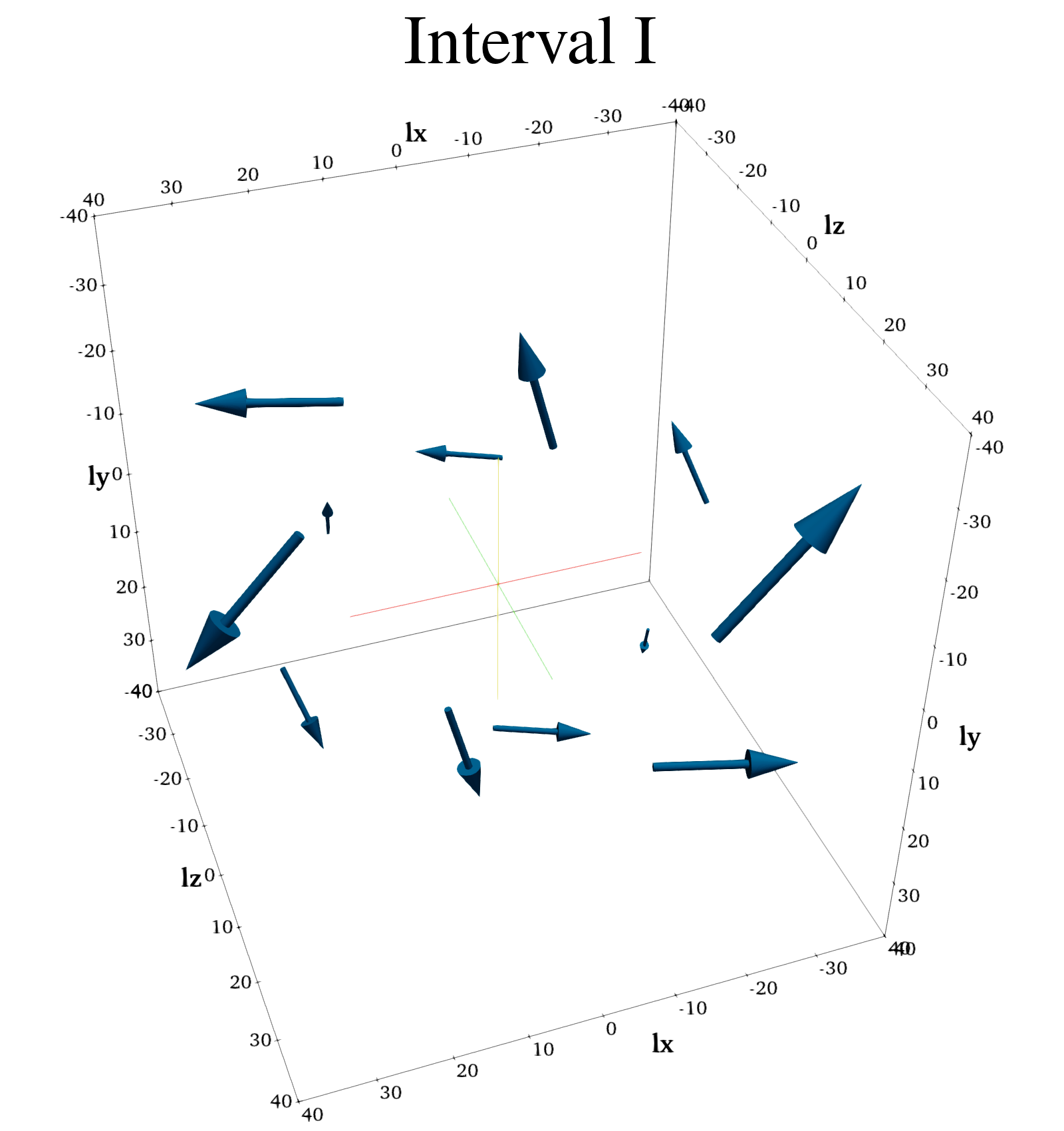}
    \includegraphics[width=0.23\textwidth]{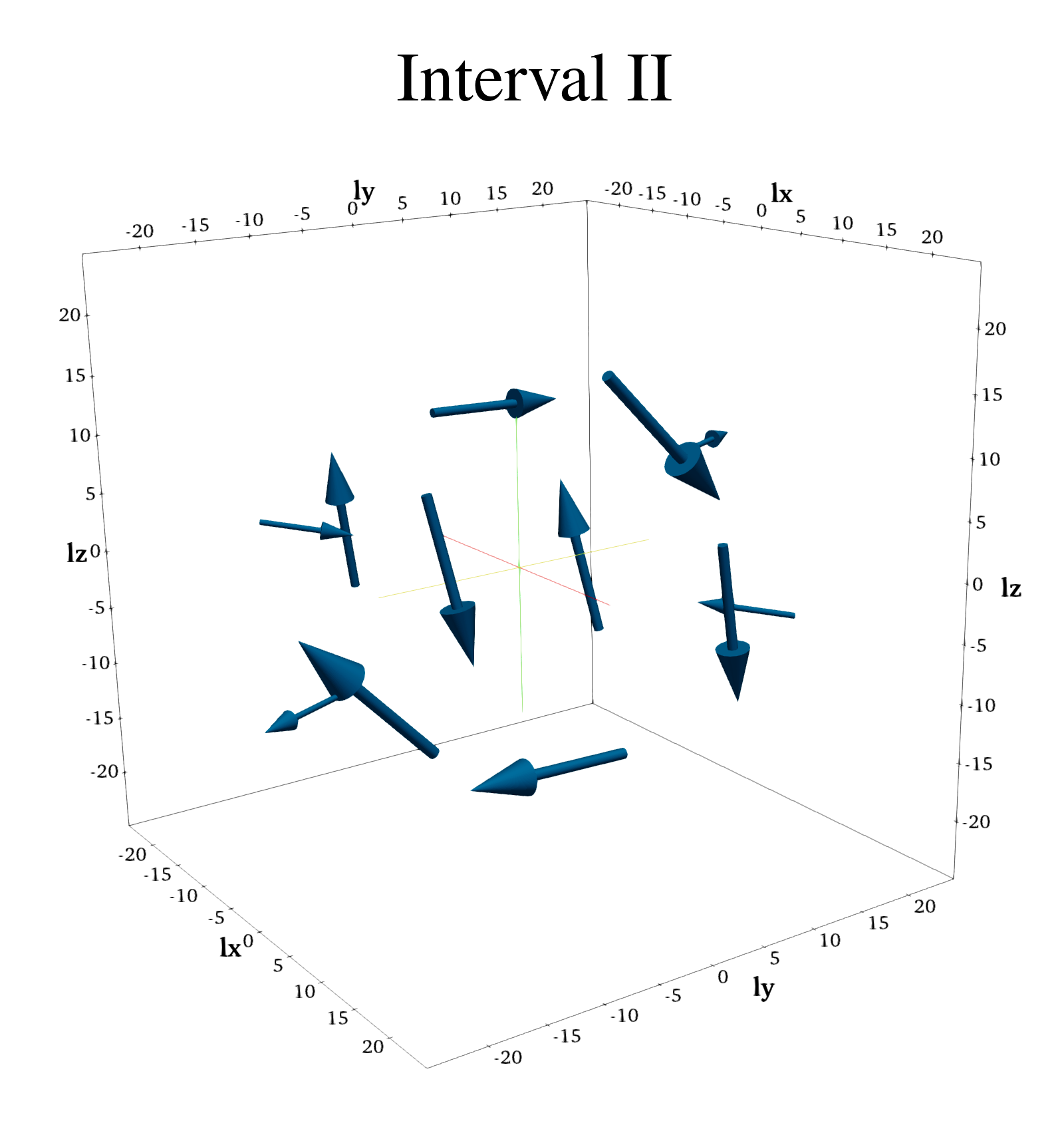}
    \includegraphics[width=0.23\textwidth]{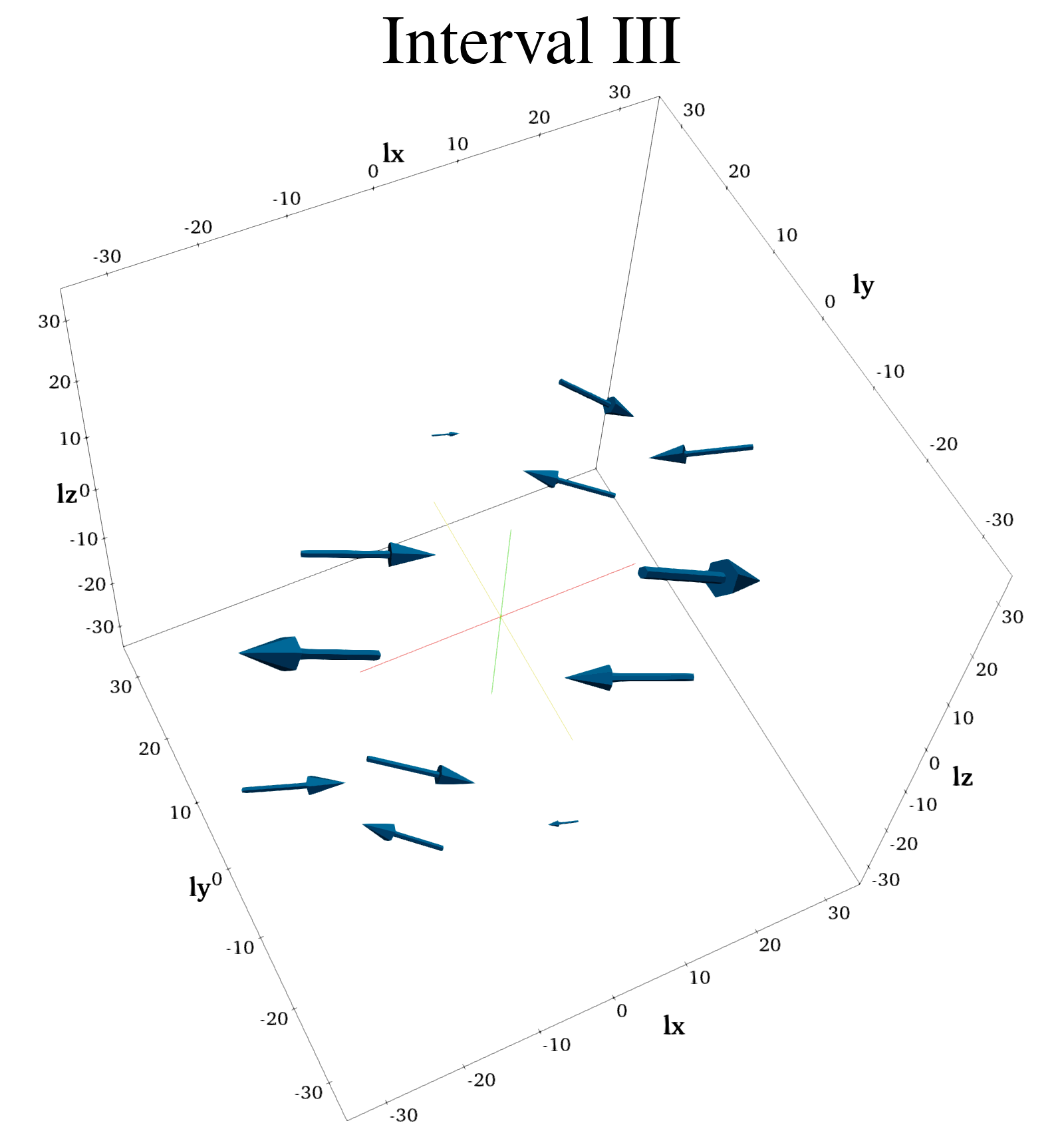}
    \includegraphics[width=0.23\textwidth]{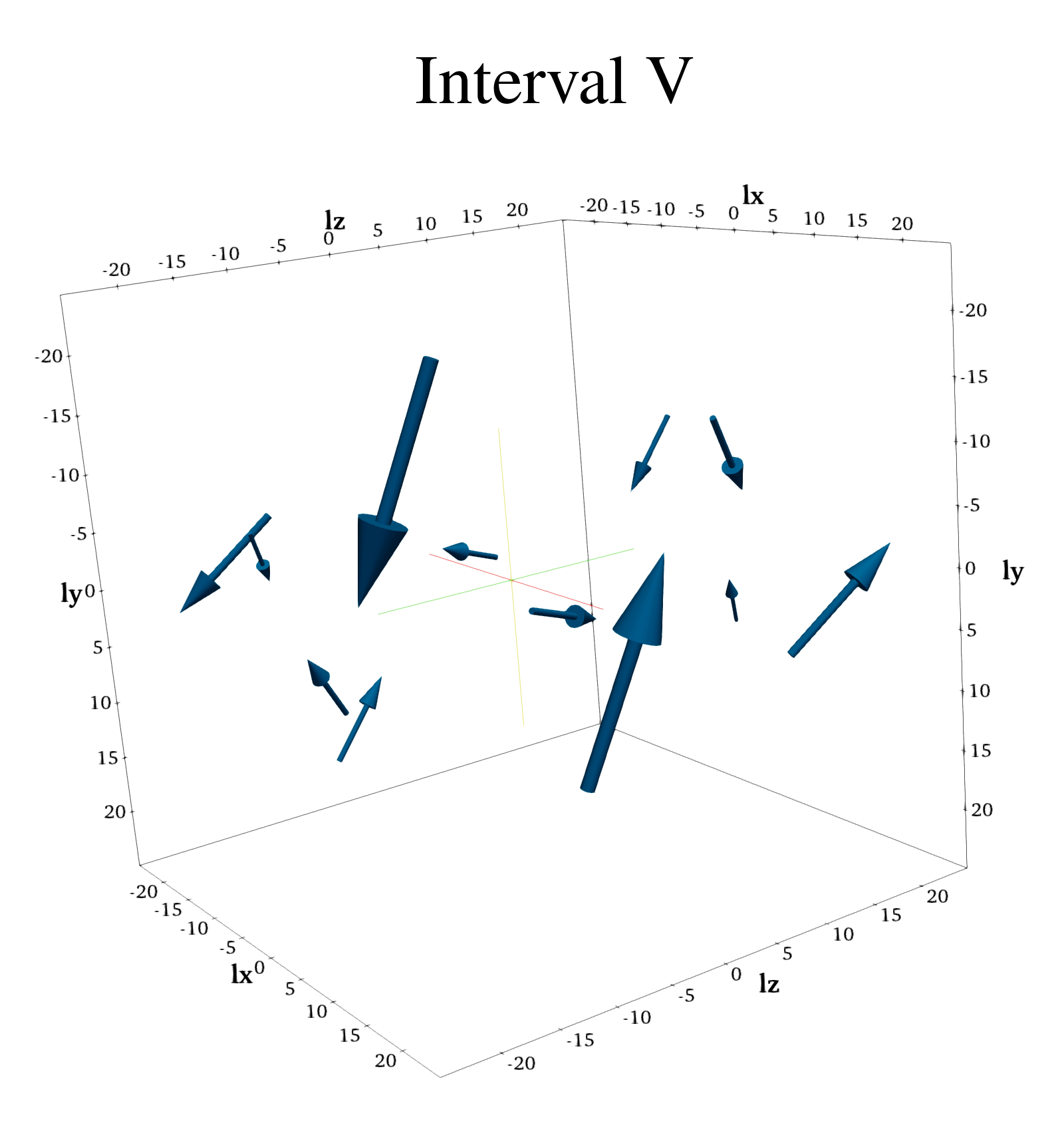}
    \caption{Yaglom flux vectors observed in the studied intervals. In all cases, the arrows (whose length is proportional to the magnitude of the vector) are larger farther away from the origin (intersection of the red, green, and yellow thin lines) of the lag space and become increasingly smaller as the origin is approached where energy is eventually dissipated. None of the intervals show the radial behavior expected in isotropic media. Interval I, differently from the others, shows $\YY$ fluxes that swirl away from the origin (positive divergence, and negative transfer rate).}
    \label{fig:yflux}
\end{figure*}

The main ingredient of Eq.~\ref{eq:yaglom} is the Yaglom flux. Historically, its magnitude has been the quantity of relevance for the determination of the energy cascade rate in space plasmas \citep{sorrisovalvo2007observation,marino2008heating,carbone2009scaling}. This is a necessary approach when using single spacecraft, but it has also been used when multispacecraft observations are available \citep{bandyopadhyay2020insitu, roy2022turbulent}. In \citet{pecora2023threedimensional_PRL}, for the first time, the vectorial nature of the Yaglom flux in a space plasma (the magnetosheath) has been observed and analyzed. Previously, it was done only in hydrodynamics (HD) experiments \citep{lamriben2011direct} and MHD simulations \citep{verdini2015anisotropy}. It was observed that, when the system is isotropic, the Yaglom flux points radially toward the origin. However, when some degree of anisotropy is induced in the system -- by rotations in HD or including a guide field in MHD -- the Yaglom flux vectors show a deflection from the radial direction. A similar behavior was observed in the magnetosheath. Here, are reported in Fig.~\ref{fig:yflux} the vector Yaglom fluxes for intervals I, II, III, and V. None of such intervals show a radial behavior of $\YY$ suggesting that the magnetosheath is a highly anisotropic environment. Interesting to notice is interval I that is associated with a negative cascade rate and, therefore, the Yaglom flux arrows point away from the origin of lag space (negative divergence).

\section{Discussion and Conclusions}
\label{sec:disclusion}

In this work, measurements of the turbulent energy cascade rate in the Earth's magnetosheath were presented as a supplement to \citet{pecora2023threedimensional_PRL}. These estimates are based on the LPDE technique initially described and tested with MHD simulations in \citep{pecora2023helioswarm}. Here, the focus is the incompressible MHD version of the vKH equation as obtained in \citep{politano1998dynamical_PP1, Politano1998vonkarman_PP2}. Additional corrections to include compressibility \citep{carbone2009scaling,banerjee2013exact,andres2017alternative}, or Hall effects have been proposed \citep{galtier2008vonkarman,banerjee2017alternative,hellinger2018vonkarman_HMHD,ferrand2019exact} separately, or together \citep{andres2018exact}. In principle, these are additive corrections to the incompressible case considered here and may include additional hypotheses (such as an isothermal equation of state).

The measurement of the energy transfer rate in space plasmas is complicated for several reasons. It is a three-dimensional, scale-dependent quantity to be measured in an anisotropic medium, with information obtainable only at a few points (spacecraft positions). The initial approach was that provided by \citep{Politano1998vonkarman_PP2} in which Yaglom's law (Eq.~\ref{eq:yaglom}) was simplified for isotropic systems and it has been widely used since \citep{macbride2005turbulence,sorrisovalvo2007observation,carbone2009scaling,roy2022turbulent}. This approach, however, necessarily assumes isotropy and the validity of Taylor hypothesis \citep{taylor1938spectrum_frozenin}.

Regarding isotropy, several works have shown that the interplanetary medium is anisotropic \citep{matthaeus1990evidence, dasso2005anisotropy, weygand2009anisotropy}. Magnetohydrodynamics simulations have shown the large variability induced by the assumption of isotropy when the energy transfer rate is estimated along different directions with respect to a preferred axis \citep{verdini2015anisotropy, wang2022strategies, jiang2023energy}. Assuming Taylor's hypothesis possibly raises entirely different issues related to the space-time correlation \citep{servidio2011time, matthaeus2016ensemble}, especially when long datasets are used -- as it is required for the third-order structure function to converge \citep{stawarz2009turbulent}.

Improvements to the isotropic approach have been developed in the past in order to obtain more refined estimates of the energy transfer rate. A combination of 2D+1D models assuming gyrotropy leads to the evaluation of the cascade rate in the directions parallel and perpendicular to a certain direction (possibly governed by the mean magnetic field) \citep{macbride2008turbulent,stawarz2009turbulent}. When adequate coverage in angles is obtainable, it is possible to solve the Yaglom law integrating over a sphere without assuming any particular geometry \citep{osman2011anisotropic}. This latter approach, however, necessarily requires datasets long enough (at least several months) to obtain such coverage and therefore it cannot be applied for the analysis of short time intervals.

An additional issue concerned the signed nature of the transfer rate which can be either positive or negative. Statistical significance is needed in order to rely on the obtained estimate instead of its absolute value \citep{stawarz2009turbulent,hadid2018compressible}.

The results presented here show how several of these issues can be surmounted using the LPDE technique \citep{pecora2023helioswarm,pecora2023threedimensional_PRL}. Depending on the number of spacecraft available (in a number greater than 3), it is possible to solve Yaglom's equation several hundred to several hundred thousand times with no underlying hypotheses for any given time interval. It is not necessary to assume the frozen-in condition since derivatives in lag space are taken among different pairs of spacecraft at the same time. No assumption on geometrical symmetries is required to solve the full vectorial equation since well-known curlometer-like algorithms are utilized \citep{dunlop1988analysis}.
Additionally, the large number of estimates, obtained for intervals of any length, provide solid statistical significance. The general behavior consists of the majority of the estimates having one sign, and few to none of the opposite sign, confirming the reliability of LPDE. For approaches for which only a single value per scale is available, it is unclear whether that is of the correct sign (or even magnitude).

This approach naturally led to the observation of the vector Yaglom flux in space plasmas for the first time. Previous HD experiments \citep{lamriben2011direct} and MHD simulations \citep{verdini2015anisotropy} have shown that the Yaglom flux deviates from the radial behavior expected for isotropic turbulence when some degree of anisotropy is induced in the system. In the magnetosheath, as presented here, such non-radial behavior is observed as well. The immediate conclusion is that this is an anisotropic environment. Speculatively, this may motivate the search for an additional equation to pair with Eq.~\ref{eq:yaglom} which involves the curl of the Yaglom flux and some associated measure of the anisotropy of the turbulent fluid.

The LPDE technique is limited to the range of scales provided by the spacecraft constellation if the Taylor hypothesis is not used. Therefore, the cascade rate that is obtained is a ``partial'' cascade rate due to the nonlinear transfer at scales not necessarily in the inertial range as it is for the present case where the MMS separations are sub-$d_i$. However, the accuracy of this estimate is such that the obtained estimates can be used as a solid lower threshold for the total energy cascade rate.  This limitation will be overcome when the new multipoint multiscale missions, HelioSwarm \citep{spence2019helioswarm,klein2023helioswarm} and Plasma Observatory \citep{retino2022particle_PO}, will be launched and proper measurements of the scale-dependent turbulent energy transfer rate will be within our reach.

\begin{acknowledgments}
This work at the University of Delaware has been
supported by NASA MMS Mission under grant number 80NSSC19K0565,
the NASA MMS GI grant (subcontract from Princeton  NNH20ZDA001N-HGIO/SUB0000517),
and a Plan for NASA EPSCoR Research Infrastructure Development (RID) in Delaware (NASA award 80NSSC22M0039).
\end{acknowledgments}





\end{document}